# Unraveling Medium-Range Order and Melting Mechanism of ZIF-4 under High Temperature

Zuhao Shi[1], Bin Liu[1], Yuanzheng Yue[1,2], Arramel Arramel[5], Neng Li[1,3,4*]

[1]State Key Laboratory of Silicate Materials for Architectures, Wuhan University of Technology, Wuhan 430070, China

[2]Department of Chemistry and Bioscience, Aalborg University, DK-9220 Aalborg, Denmark

[3]State Key Laboratory of Advanced Technology for Float Glass Technology, CNBM Bengbu Design & Research Institute for Glass Industry Co., Ltd, Bengbu 233010, China

[4]State Center for International Cooperation on Designer Low-Carbon & Environmental Materials (CDLCEM), School of Materials Science and Engineering, Zhengzhou University, Zhengzhou 450001, Henan, China

[5]Nano Center Indonesia, Jalan Raya PUSPIPTEK, South Tangerang, Banten 15314, Indonesia

*Correspondence: lineng@whut.edu.cn

**ABSTRACT:** Glass formation in Zeolitic Imidazolate Frameworks (ZIFs) has garnered significant attention in the field of Metal-Organic Frameworks (MOFs) in recent years. Numerous works have been conducted to investigate the microscopic mechanisms involved in the melting-quenching process of ZIFs. Understanding the density variations that occur during the melting process of ZIFs is crucial for comprehending the origins of glass formation. However, conducting large-scale simulations has been challenging due to limitations in computational resources. In this work, we utilized deep learning methods to accurately construct a potential function that describes the atomic-scale melting behavior of Zeolitic Imidazolate Framework-4 (ZIF-4). The results revealed the spatial heterogeneity associated with the formation of low-density phases during the melting process of ZIF-4. This work discusses the advantages and limitations of applying deep learning simulation methods to complex structures like ZIFs, providing valuable insights for the development of machine learning approaches in designing Metal-Organic Framework glasses.

**Introduction**

The metal–organic frameworks (MOFs) have attracted significant attention in recent years,[1] particularly in relation to their response to changes in temperature and pressure. The emergence of MOF glasses has considerably expanded the research boundaries within the field of MOFs.[2] Both MOF crystals and glasses demonstrate a great potential in gas absorptions,[3,4] photoelectrics[5,6] and other fields.[7-9] As MOF glass research is at an early stage, there is an urgent need to unravel the structure-property correlation (SPC) of MOF glasses that lack of the ordered structure at various length scales.[10]

Understanding how the atomic arrangement within MOF glasses relates to their unique properties is crucial for advancing their development and unlocking their full potential. Besides experimental methods, atomistic simulation is also a powerful tool to unveil the SPC of MOF glasses. Several simulation studies employing both first-principles methods and molecular dynamics have yielded insights into microstructure of MOFs glasses.[11-14] Such studies have played a crucial role in advancing our understanding of MOF glasses and guiding further experimental efforts.

In the amorphization of MOF, the first-principles molecular dynamics (FPMD) reveals the exchange behavior of organic ligands in MOF glasses.[11,15] The spatial position resistance of ligands has a strong influence on the glass-forming ability.[16] The reactive forcefield methods have also contributed insight into the medium-range structure evolution.[17-20] The optical properties of MOF glasses have been linked to their electronic structures by utilizing on large-scale amorphous MOF model.[21,22]

The highly accurate first-principles simulations are computationally expensive when dealing with large-scale systems, in which the spatial and time scales are often limited. In contrast, the classical molecular dynamic simulation is strongly dependent on the reliability of the potential function to flexible Zeolitic Imidazolate Frameworks (ZIFs), especially for the melting process of several ZIFs.

The ReaxFF method is generally recognized for striking a good balance between accuracy and efficiency in simulations.[17] This method combines elements of classical force fields and quantum mechanical calculations, which allows for the treatment of chemical reactions and bonds breaking/formation. It could make a great contribution to revealing the SPC in glassy ZIFs.[23-25] The applicability of a generic model, developed to use a large number of static structures of ZIFs, needs to be verified for specific phase transition processes of ZIFs under certain temperature and pressure conditions.[26]

Recently, the machine learning methods have made substantial progress in the development of simulations.[27,28] Among the various machine learning methods, the combination of deep learning method and molecular dynamics has accelerated the exploration of new structures.[29] For complex chemical structures, such as high-entropy alloys and large protein molecules, the deep-learning potential molecular dynamic (DPMD) has revealed their SPC in larger spatial and time scales.[30-33] Recent studies have shown that DPMD, when trained with a well-curated dataset, can yield accurate predictions for dynamic properties like viscosity of alloys.[34] DPMD combines elements of machine learning and molecular dynamics to capture the complex dynamics of materials.

In this work, a deep-learning potential function was utilized to simulate the melting process of three representative ZIFs. Thanks to the latest version of the open-source DeepMD-kit,[35] the deep potential molecular dynamics (DPMD) simulations provided high accuracy comparable to density functional theory (DFT) results. Moreover, DPMD simulations achieved a significant breakthrough in terms of spatial and temporal scales, with an increase of three orders of magnitude. The highly efficient DPMD simulations revealed density variations associated with phase changes during the melting of ZIFs. Additionally, we explored the correlation between the heating velocity and the phase-transition point through the simulations. Finally, we discussed the challenges that DPMD currently faces in the study of amorphous ZIFs.

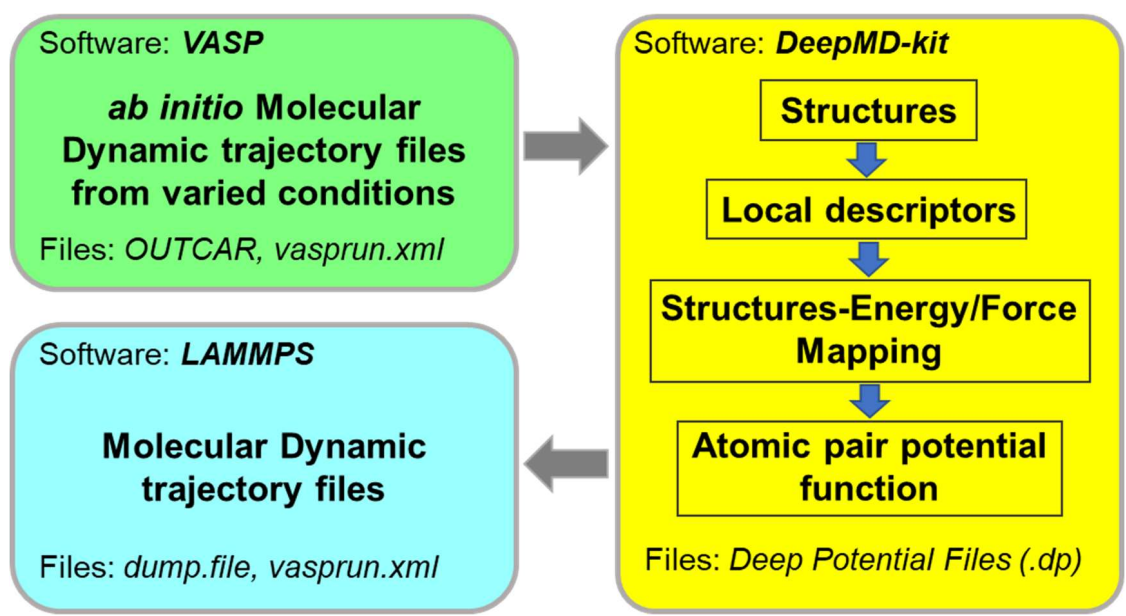

**Fig. 1. Schematic of deep learning accelerated molecular dynamics.** The structural and energetic relationships obtained from first-principles calculations will be employed in the construction of deep learning potentials. The atomic structures will be decomposed into local environments, and the mapping relationships between the structure, energy, forces, and virial quantities will be established using deep learning techniques. The potential functions constructed through deep learning will be utilized for simulations at larger time and spatial scales in LAMMPS.

**Results and Discussions**

In this work, the DeepMD-kit was used to construct the neural network (NN) potential, since this code has been continuously improved in recent years.[35] The entire workflow of this study is demonstrated in Fig. 1, where the FPMD that generated the training dataset is combined with the MD simulation using the deep-potential function. The detailed description of parameters in this workflow is depicted in Methods. A major advantage of DPMD is its ability to perform efficient molecular dynamics simulations with DFT accuracy. Using ZIF-4 as an example, Fig. 2 depicts the comparison of model size and the simulation performance of training data (FPMD) and DPMD. Compared to ab initio methods, DPMD simulations offer a significant advantage in terms of both spatial scale and time scales of the simulation. Models with approximately 8000 atoms can be employed to perform dynamical simulations at speeds of up to 0.5 ns/day by using deep learning potentials.

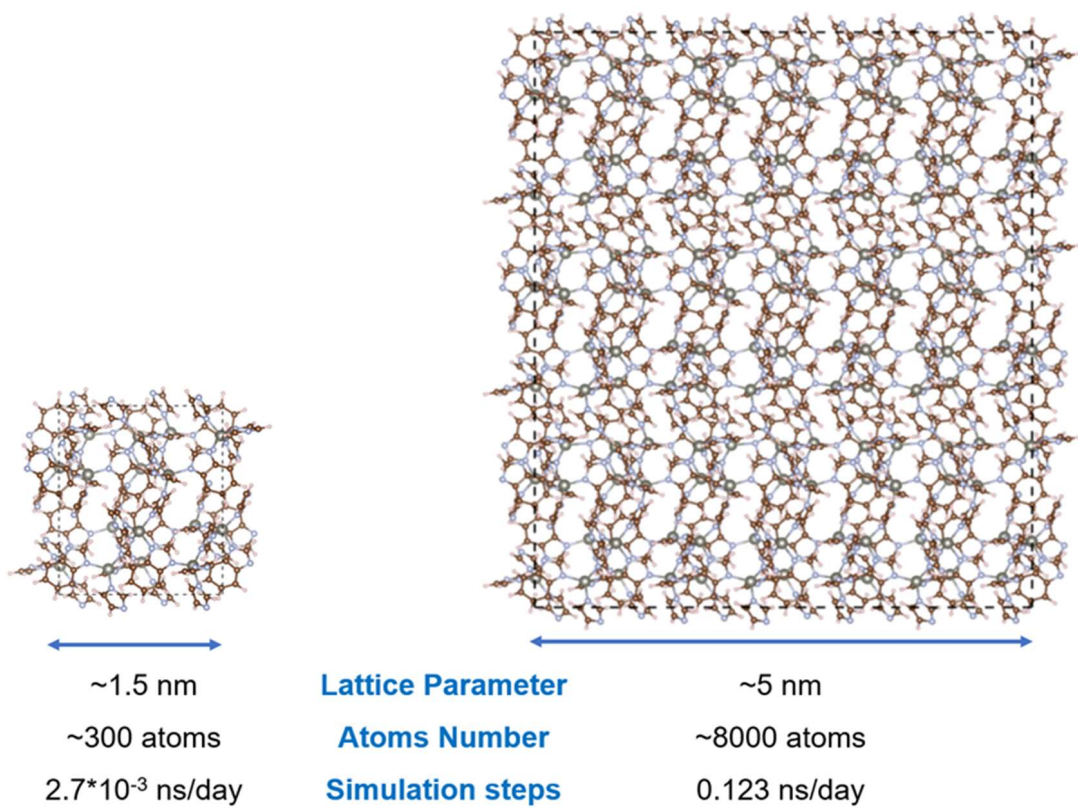

| ~1.5 nm | **Lattice Parameter** | ~5 nm |
| ~300 atoms | **Atoms Number** | ~8000 atoms |
| $2.7*10^{-3}$ ns/day | **Simulation steps** | 0.123 ns/day |

**Fig. 2. Schematic of simulation structure.** (Left) the ZIF-4 structural model used in the training set (FPMD). (Right) the supercell structure used in DPMD. Considering the model size and computational speed, the DPMD simulation is $10^3$ times more efficient than the FPMD.

In terms of the size of the simulated system, the structure undergoing DPMD in this work has lattice parameters exceeding 5 nm, meaning that the simulation involves a system of considerable size. Compared to the models used for ab initio simulations where the lattice parameters are only about 1.5 nm, DPMD is certainly more suitable for a statistical description of the local structure over the medium range (5 ~ 10 Å). It can be stated that DPMD is 5000 times more efficient in operation than FPMD for the same size system. In terms of computational speed and the size of the simulated system, DPMD is more suitable than the ab initio method for studying the evolution of structures involving meso-stable states, considering both computational speed and the size of the simulated system. It is known that most of the DFT calculations were performed under the ground states consideration or at room temperature for structural optimization, such as the QMOF database.[36] In our initial simulations, referred to as DPMD-Test, a restricted training set consisting of 20000 samples was utilized. These samples were extracted from trajectory files of ZIF-4 structures at a temperature of 300 K. We aim to access the generalizability of the deep learning potential function, which was carried out under a training of structured sample at specific temperature, by simulating sample deformations over a broader temperature range.

The optimized ZIF-4 structures were heated from 300 K to 2100 K at a heating velocity of 20 K/ps. As a result, the equilibrium structure of DPMD-Test is determined at 1200 K and fairly compared to the FPMD simulation results. Fig. S1 shows the radial distribution function of the equilibrium structure in DPMD-Test at 1200 K. The total radial distribution function displays a peak in the range of 0-10 Å. By examining Fig. S2, this peak should be associated with the Zn-Zn atomic pairs. Upon examining the radial distribution functions of the different atomic pairs in Fig. S2, it is found that both DPMD-Test and VASP calculations exhibit a consistent behavior for the atomic pairs with chemical interactions such as the metal-ligand bonding and the intermolecular bonding in ligands. In contrast, DPMD-Test was poorly performed to describe the medium-range structures. The results of DPMD-Test indicate that utilizing structures from a single temperature as the training seeds lead to a poor performance in describing the medium-range structures. To solve this problem, we constructed a larger training set by incorporating structural information at different temperatures, and thereby improved the performance and accuracy of the model.

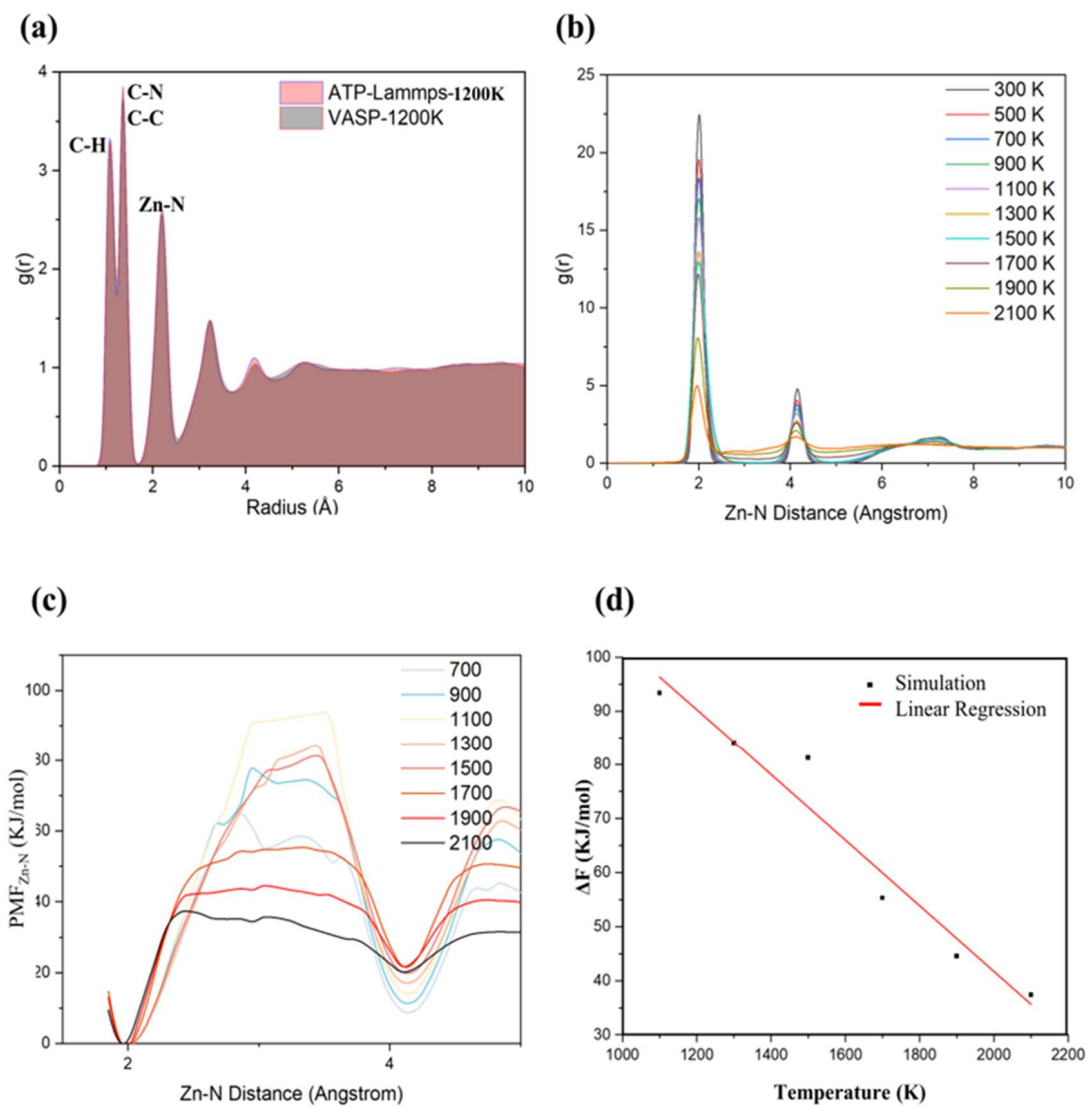

**Fig. 3. Static characterization of generative structures under deep learning training potential functions.** (a) The total radial distribution function of DPMD-20K and FPMD simulations and (b) Zn-N pair distribution function of DPMD-20K. (c) The PMF of Zn-N and (d) linear regression of free energy change (ΔF) at varied temperature in DPMD-20K.

To increase the applicability of the deep learning potential function at different temperatures, the training sets were extended by including all sets mentioned in Methods part. In this attempt, the boundary conditions of more than 200,000 configurations of ZIF-4 and ZIF-62 FPMD trajectories is considered. We refer to this new potential function as ATP-D3 (All Temperature Potential). The predicted performance is shown in Fig. S3. The root mean-square error of energy, force, and virial for three ZIFs are listed in Table 1. Taking into account the DPMD training results to be implemented for other complex systems, such as high-entropy alloys or heterogeneous catalytic interfaces,[37] we believe that the accuracy of the model is sufficient for simulating the structural evolution during the subsequent melting process.

**Table 1.** The error of DPMD for varied dataset of three ZIFs (ZIF-4, ZIF-62 and ZIF-8). The italic letter means relative error.

| Datasets | Energy (meV/atom) | Force (eV/Å) | Virial (meV/atom) |
| --- | --- | --- | --- |
| ZIF-4 (300 K) | 3.61 (0.10%) | 1.14 | 20.5 |
| ZIF-4 (2500 K) | 3.37 (0.09%) | 1.25 | 13.8 |
| ZIF-62 (300 K) | 4.41 (0.11%) | 1.20 | 25.2 |
| ZIF-62 (2500 K) | 6.30 (0.15%) | 1.30 | 15.7 |
| ZIF-8 (300 K) | 37.1 (1.12%) | 2.02 | 16.1 |

In addition to the ZIF-4 and ZIF-62 structures included in the training set, we tested the structure of ZIF-8 at 300 K to examine the generalization of the potential function. The results demonstrate that the deviation of energy prediction for ZIF-8 at 300 K is over 1%, which is approximately ten times higher than ZIF-4 and ZIF-62. Due to the significant differences in topology and functional group species across ZIF-4, ZIF-62, and ZIF-8, the structure-energy relationships constructed according to the XIF-4 structural order covariates are not effective to accurately describing the energy of ZIF-8. This suggests that a deep-learning potential functions can be applicable for the various ZIFs, such as ReaxFF,[17] then a broader phase space input should contain different ligands and topology information. To evaluate the performance of the ATP, we performed a series of simulations at different heating velocity. We first performed the ZIF-4 heating process at 20 K/ps, since this heating velocity is very close to the setting in typical FPMD (15-20 K/ps).[11,15,16] We therefore conducted a statistical comparison of the distribution of the static structures. Fig. 3(a) shows the comparison in the radial distribution function between the DPMD-20K and FPMD structures at 1200 K. Compared to DPMD-Test shown in Fig. S1, it can be observed that the structures of DPMD-20K and FPMD are almost identical in their radial distribution functions.

Table 2 The comparison in thermodynamic values between DPMD and FPMD. $\Delta U$, $\Delta S$ and $\Delta F$ represent the changes of internal energy, entropy and free energy respectively, following a van 't Hoff law $\Delta F = \Delta U - T*\Delta S$.

| Ref | $\Delta U$ (kJ/mol) | $\Delta S$ (J/mol/K) | $\Delta F$ (840K) (kJ/mol) | $\Delta F$ (1800K) (kJ/mol) |
|---|---|---|---|---|
| Gaillac *et. al* (Ref 12) | 127 | 37 | 95.9 | 60.4 |
| Shi, et. al (Ref 17) | 164.3 | 44.1 | 127.3 | 84.9 |
| DPMD-20K (This work) | 163.1 | 60.7 | 112.1 | 53.8 |

The evolution of the structure with temperature were. Fig. 3(b) shows the Zn-N pair distribution function (PRDF) of DPMD-20K at different temperatures. The intersection between the first peak and the second peak indicates that partial melting can occur when the temperature rises above 1100 K. We calculate the potential of mean force (PMF) of Zn-N pairs by the PRDF. Fig. 3(c) and 3(d) exhibit the temperature dependence of the activation free energy, while Table 2 shows the comparison between DPMD-20K and previous DFT calculations. The error in the structure evolution is smaller in the free energy, while DPMD has a higher value in the change of entropy. According to the PMF fitting formula, the linear fitting formula corresponding to DPMD-20K has a higher slope than the previous DFT results, indicating the energy of DPMD-20K decreases faster than its counterparts.

Due to the high computational efficiency of DPMD, we were able to simulate the heating process of three ZIFs at a slower heating velocity than typically FPMD process. We chose ZIF-4 to explore the influence of heating velocity on structures. To ensure that ZIF-4 structures reach equilibrium at each temperature, we used a segmented equilibrium approach for heating ZIF-4 at 20 K/ps. In addition, the heating process of ZIF-4 at 4 K/ps and 1 K/ps used a one-step heating method to observe the continuous change of ZIF-4 structures.

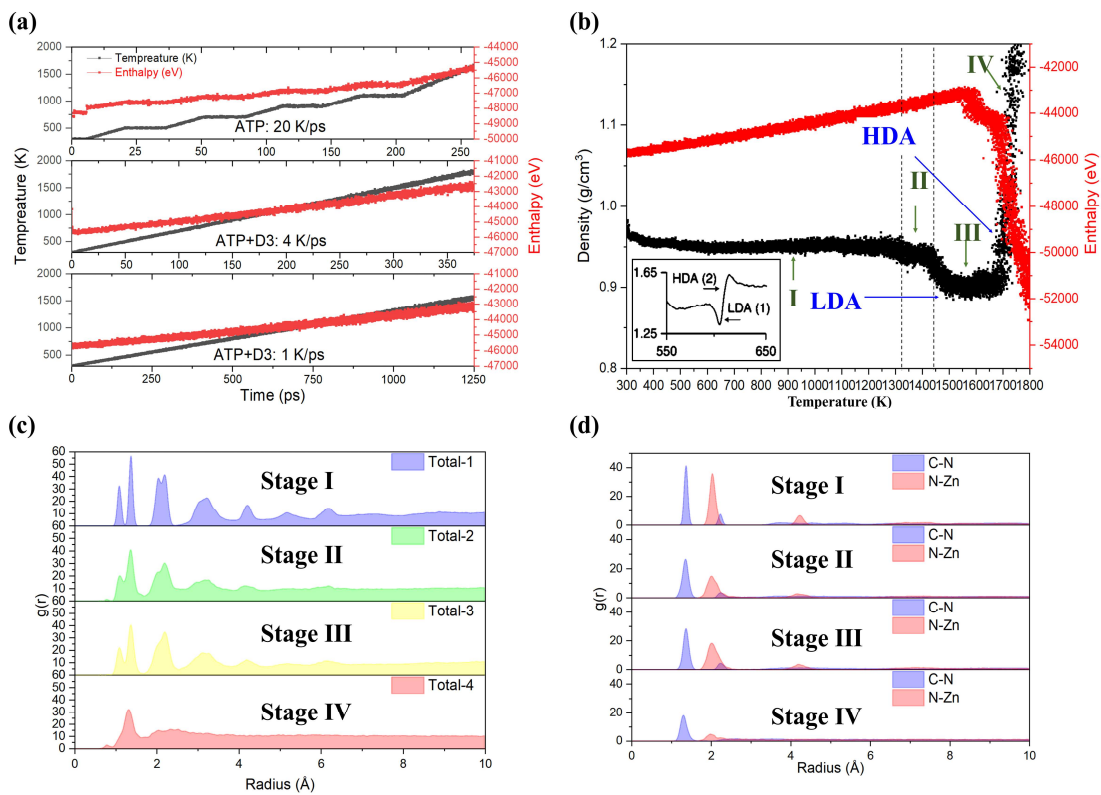

**Fig. 4. Medium-range structural changes of ZIF-4 in DPMD.** (a) The enthalpy changes of melting process at varied heating rate. (b) Density and enthalpy changes during the heating of ZIF-4 at 1K/ps (DPMD-1K). The red and black curves show the change in enthalpy and density, respectively. The inset diagram is the isobaric heat capacity ($C_p$) of ZIF-4 upon heating and melting, exothermic collapse to the LDA phase (1) which is closely followed by (2) endothermic formation of the HDA phase, Reprinted with permission from ref [38]. Copyright 2015 Springer Nature Limited under Creative Commons license CC BY 4.0. https://creativecommons.org/licenses/by/4.0/. (c) Total radial distribution function and (d) pair radial distribution function of ZIF-4 at different stages of the heating process of DPMD-1K.

The energetic variation as a function of temperature at different heating velocity is shown in Fig. 4(a). Fig. 4(b) displays the variation of density of ZIF-4 with the temperatures. It turns out that the density of ZIF-4 is in dynamic equilibrium over the wide temperature range from 300 K to around 1300 K. The density of ZIF-4 within the aforementioned temperature range is about 0.95 g/cm$^3$, which is much lower than that of a real anhydrous ZIF-4 structure. We postulate that the DFT data used for the training is not fully describe the density variation of ZIF-4. In the training data, the volume was fixed to be a constant in NVT ensemble, and hence the change of the long-range structure can be neglected. This has less impact on volume change for the scale of the FPMD, however, when this structure-energy relationship is reproduced at larger scales, it causes bias in the density. As the temperature over 1300 K, the density of ZIF-4 undergoes a non-negligible reduction. As seen in the structure diagrams in Fig. S4(a) and 4(b), ZIF-4 structure does not undergo a significant transformation at this point, and the decrease in density is attributed to the thermal expansion of the lattice. With further incremental changes in temperature, the ZIF-4 structure shows a significant low-density phase in the range of 1400-1550 K. The formation of low-density ZIF-4 has been reported in experiments, where the low-density amorphous phase (LDA) appears before the high-density amorphous phase (HDA) at temperatures around 600K.[38-40] However, in our simulations, the temperature at which the low-density phase emerges is significantly higher than the experimental value. One possible reason for this discrepancy is that the simulated heating rate and system size are much smaller than those in the experimental samples, resulting in inadequate relaxation of the ZIF-4 crystal structure.

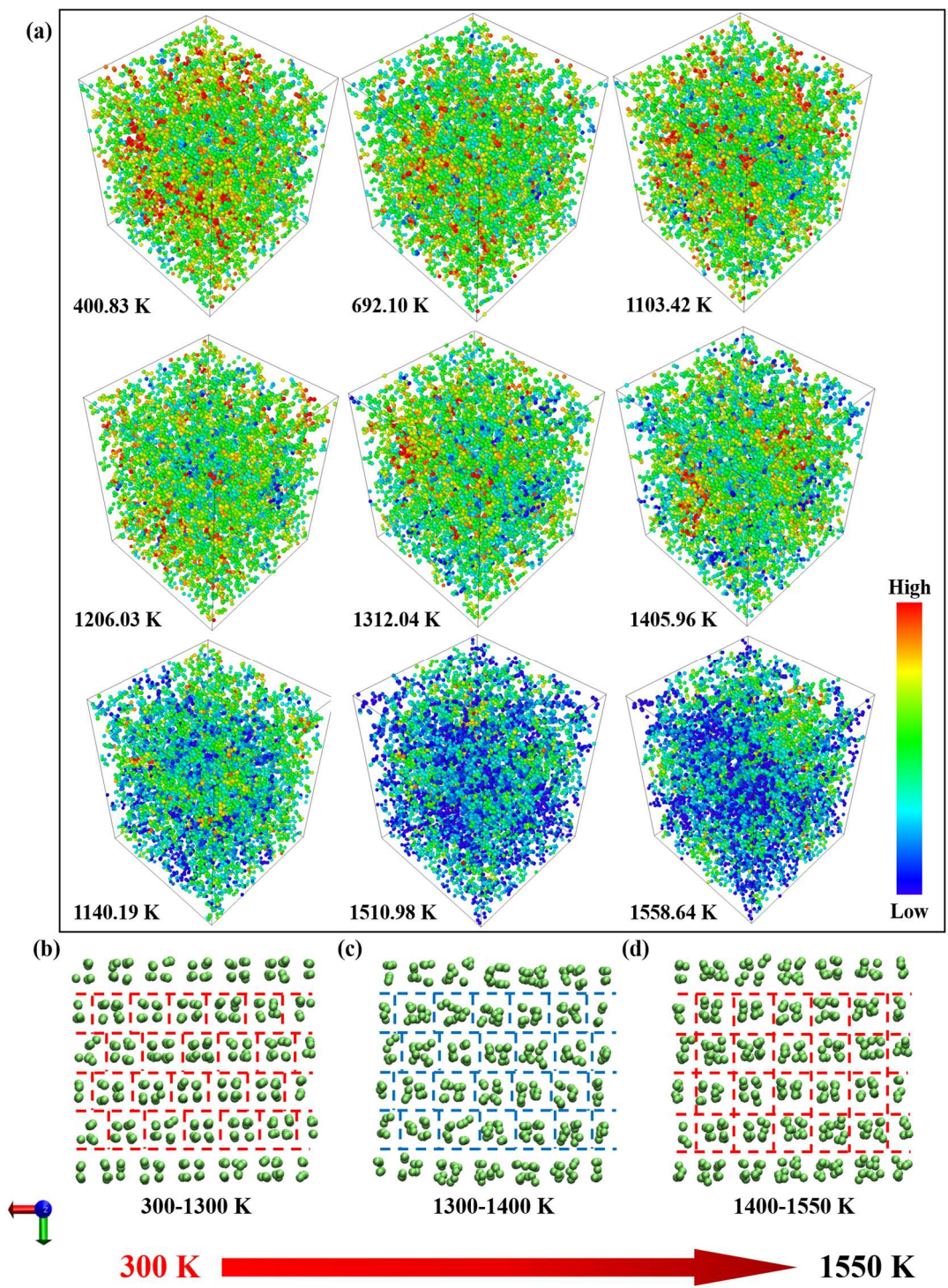

**Fig. 5. Spatial heterogeneity in the distribution of densities during ZIF-4 heating.** (a). The variation of the density distribution of the ZIF-4 structure with temperature. The color bar indicates the density from low (blue) to high (red); (b)-(d). Evolution of mid-range structure in the melting process of ZIF4 in DPMD-1K at varied temperature. Only Zn atoms (green balls) are preserved and displayed (along the z-axis direction).

The manifestation of the radial distribution function for the different phases of the structure is shown in Fig. 4(c), where the intersection between the second and third peaks in the second and third stages reflects the partial melting of ZIF-4. The breaking of metal-organic ligand bonds (Zn-N pairs) is the cause of partial melting, as demonstrated in Fig 4(d). With further increasing the temperature, ZIF-4 becomes denser. From the structure above 1550 K in Fig. S4(c), we can obtain the high-density crystalline phase ZIF-zni.[10,41] As the temperature increases beyond 1700 K, ZIF-4 undergoes melting and subsequent decomposition, and non-physical meaning structures, such as the clusters formation of H atoms. The formation of the LD phase of ZIF-4 is only observed when the heating velocity is low enough (1K/ps). For the typical software based on the DFT method, a significant computational resource is required to perform molecular dynamics at the mentioned

To further investigate the formation process of low-density phase, we visualize the variation of the density distribution of the ZIF-4 structures with temperature in DPMD-1K (Fig. 5). The local density is defined by the number of atoms in the 5 Å * 5 Å * 5 Å box, with green and red colors indicating low and high local atomic densities, respectively. The statistical distribution of the medium-range structure of ZIF-4 is displayed in Fig. 5(b)-(d) in different temperature ranges, corresponding to the heating-up stage of 300-1300 K, the pre-melting stage of 1300-1400 K and the low-density melt stage of 1400-1550 K. We can observe a uniform high-density local phase distribution in the ZIF-4 structure during the heating-up stage. This is due to the accelerated local atomic motion caused by the increasing of the temperature. As the temperature increases, the lattice of ZIF-4 undergoes considerably expansion during the pre-melting stage, leading to a decrease in high-density regions within the structure. At the temperature surpasses 1400 K, an increased number of low-density regions emerge within the structure of ZIF-4. However, when the temperature is higher than 1550 K, the high-density regions reappear.

It is important to note that the distribution of different density local phases within the ZIF-4 structure is spatially heterogeneous. To further discuss the generation of this spatial heterogeneity, we plot the root-mean-square displacement of different atoms as a function of temperature (Fig. S5). Several constituents of the organic ligand functional groups, represented by N, C and H atoms, are diffused faster at lower temperatures compared to the metal node Zn atoms. These verify the short-range exchange effect during the melting of ZIF-4 as revealed by the previous simulation work.[11,15,16] As the temperature increases, the organic functional group progressively detaches from its original metal node. In the short-range structure, this results in the creation of undercoordination $[ZnN_x]$ (x<4). The spatial constraints are reduced due to the decreased average Zn atoms coordination in the medium-range structure, resulting in a higher degree of freedom for Zn nodes. As depicted in Fig. 5(b)-(d), Zn atoms with smaller constraints are rearranged in the medium-range structure. This results in the formation of a low-density melt phase.

The Lindemann ratio ($\Delta$), calculated using the full width at half maximum (FWHM) of the radial distribution function, has been demonstrated as a reliable structural order parameter for describing the phase transition structures. The FWHM is depicted in Fig.

6 (a). We further explore the short- and medium-range structure transition of ZIF-4 by examining the FWHM of the Zn-N and Zn-Zn pairs. The FWHM shows an increase as the temperature rises. This observation is supported by the variations in the pair distribution function shown in Fig. 3(b), reflecting a softening of the network structure.

The FWHM of the Zn-N and Zn-Zn pairs is consistent at 300-700 K, indicating that ZIF-4 is in a stable crystalline state at the early stage of the heating-up stage. With an increase in temperature, the FHWM of the Zn-N pair increases faster than that of the Zn-Zn pair. This phenomenon demonstrates that the metal-ligand bonds in ZIF-4 is more sensitive to temperature, i.e., ligand detachment from the metal node occurs. It is noteworthy that the FHWM of the Zn-N pair does not change within 1200-1400 K, while the FHWM of the Zn-Zn pair decreases. Since the FHWMs for different radial distribution functions were obtained by averaging over 1000 sampled structures, the resulting statistical error can be considered small. One possible reason is that the medium-range structural transition occurs in ZIF-4 during the pre-melting phase, as demonstrated in Fig. 5(c) and (d). During this transition, the short-range structure of ZIF-4 reaches equilibrium through ligand exchange interactions. This involves the dynamic equilibrium of breaking and reclosing between the organic ligand and different metal nodes within the structure.

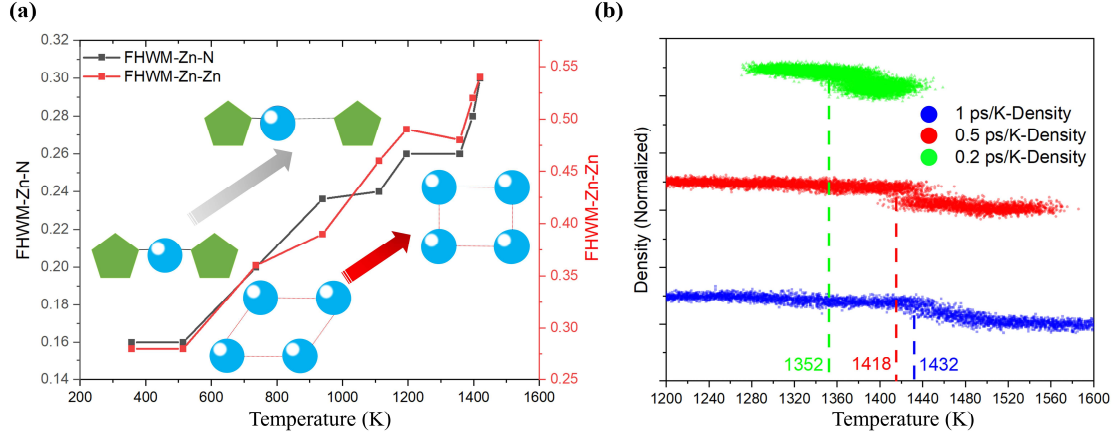

**Fig. 6. Order parameters variation of the ZIF-4 heating process.** (a) FHWM of Zn-N and Zn-Zn pairs at varied temperature; (b) Density variation of ZIF-4 at 1 K/ps, 0.5 K/ps and 0.2 K/ps heating velocity.

The temperature point at which the density decreases significantly is noted as the Transition Temperature ($T_t$). As in Fig. 6 (b), it is show that $T_t$ = 1432 K in DPMD-1K. As the heating velocity decreases, $T_t$ decreases to 1418 K for ZIF-4 at 0.5 K/ps and further decreases to 1352 K at 0.2 K/ps. The decrease in $T_t$ with decreased heating velocity is ascribed to the extension of the relaxation time of ZIF-4 at certain temperature. The experimental heating velocity for melting ZIF-4 is usually 10-20 K/min,[3,41-43] which makes it difficult to reproduce the condition of the simulation. However, the relationship between $T_t$ and heating velocity demonstrated in DPMD offers the possibility of reproducing the experimental conditions.

The generalizability of deep learning potential functions trained on specific classes of ZIFs poses a significant challenge due to variations in topology and constituent elements among different ZIFs. The extensive preparation of a training set highlights the need for further improvements in the generalizability of machine learning potential functions. To address this challenge, we propose incorporating a priori knowledge from experimental data into the hyperparameters used in constructing potential function. This approach emphasizes the importance of going beyond sole reliance on raw simulation data.

Although this research is still in its early stages, we would like to highlight two noteworthy points that researchers should consider when building upon this work. Firstly, the DeePMD-kit demonstrates remarkable efficiency in parallel computing on large-scale computers, making it a powerful tool for high-performance computing. Furthermore, it is worth mentioning that part of the deep potential function training described in this study was successfully conducted on a modest home computer. This indicates that powerful computing resources are not necessarily limited to supercomputers. Researchers can take advantage of the capability to train with historical data and subsequently employ DPMD simulations on platforms like LAMMPS. This development alleviates concerns about outdated research equipment and empowers researchers engaged in cross-scale simulations to focus on their research content. However, it is important to note that the current DPMD method represents a transitional approach and may appear as a black box to certain users. While DPMD enables researchers to extend their research systems in terms of time and space scales, its generalization to similar structures is still insufficient compared to traditional many-body potential functions. Researchers with a background in first principles or quantum chemical simulations can benefit from the DPMD method to expand their investigations and effectively utilize their historical data.

**Discussions and Conclusions**

In this work, we explored the application of deep learning accelerated molecular dynamic in amorphization of ZIF-4. By exploring the simulation conditions, the main conclusions can be drawn from this work:

(1) The potential function trained using the outputs of FPMD offers high prediction accuracy within the wide temperature range. This makes it suitable for conducting simulations of same structure on larger time and spatial scales.

(2) DPMD simulations at different heating rates reveal the phase transition in the pre-melting phase, which is not observed in FPMD.

(3) The different transition temperature, at which the liquid phase appears at different heating rates, are very useful for achieving accurate simulations of experimental phenomena.

(4) The low-density state of ZIF-4 appears before melting. The ligands-induced exchange behavior is believed to be the resource of low-density state.

While the DPMD approach provides insights into the melting behavior of ZIF-4 on a larger scale, it is essential to acknowledge certain limitations or shortcomings associated with this methodology. The first is the concern about training data. As the raw FPMD is processed in the NVT ensemble, it would deviate from the true density of amorphous ZIF. As a consequence, there may be a discrepancy between the density predicted by DPMD simulations and the experimental density. The two values might not align well or be accurately fitted to each other. Another limitation associated with the method concerns the training efficiency of deep learning.

**Models and Approaches**

**ZIFs models**

The FPMD outputs of ZIF-4 and ZIF-62 in our previous study were chosen to be the training data.[16] The initial ZIF-4, ZIF-62 model used in FPMD was obtained from the CSD database.[44] In the FPMD, all the solution molecules were removed from the cell and a dry cell was obtained. The parameters description of the ZIFs structures can be found in our previous work. In this work, this model was employed to generate sufficient configurations for the training of deep potential.

In the DPMD process, the 3*3*3 supercell of ZIF-4 was built based on relaxed unit cell. The supercell of ZIF-4 contains 7344 atoms with no symmetry, and the cell parameters a = 46.19 Å, b = 45.92 Å, c = 55.28 Å, and $\alpha = \beta = \gamma = 90°$.

**First Principles Molecular Dynamic (FPMD)**

To ensure the adequacy of the training data, some of the first principles data used in this study were from a previous work.[16] All geometric structures in this study were simulated by using Vienna Ab initio Simulation Package (VASP) in projector augmented wave (PAW) method,[45,46] and the exchange-correlation energy was evaluated in the GGA-PBE potential with the DFT-D3 method.[47,48] We chose the single K-point at the zone center in the K-points setup, which was sufficient for such a large structure model with nearly three hundred atoms. The energy cut-off is set as 400 eV and the criterions for electronic and force convergence are set at $10^{-5}$ eV and $2 \times 10^{-3}$ eV Å$^{-1}$, respectively. All FPMD simulations were performed using the canonical ensemble (NVT) with the temperatures set at be 300 K, 600 K, 900 K, 1200 K, 2000 K and 2500 K. Each simulation run for a minimum of 15 ps with the total simulation time at each temperature to ensure the equilibrium of the system. The temperature was regulated using the Nose-Hoover thermostat and a timestep of 0.5 fs was employed.[49]

Table 3 The parameters of varied heating procedures

| Sets | Starting Temperature (K) | Ending Temperature (K) | Heating Velocity (K/ps) | Steps (×10³) |
|---|---|---|---|---|
| **DPMD-Test** | 300 | 2100 | 20.0 | 220 |
| **DPMD-20K** | 300 | 2100 | 20.0 | 220 |
| **DPMD-4K** | 300 | 2100 | 4.0 | 940 |
| **DPMD-1K** | 300 | 2100 | 1.0 | 3640 |
| **DPMD-0.5K** | 1200 | 1900 | 0.5 | 2820 |
| **DPMD-0.1K** | 1300 | 1600 | 0.2 | 6030 |

To expand the phase space of flexible ZIFs structures, additional molecular dynamic simulations and geometry optimization of ZIF-4 were conducted. We additionally performed the isobaric-isothermal (NPT) simulations at 300 K and 2000 K under the Langevin thermostat and zero pressure. Simultaneously, a series of structural optimizations were performed by varying the lattice constants, ranging from a -5% to +5% change. Through NPT simulations and structural optimizations, we generated a total of 50,620 structural trajectory files with flexible lattice at 0 K,

**Scheme of deep-learning potential**

The training sets included nine sets corresponds to structures at varied temperatures

(300 K, 600 K, 900 K, 1200 K, 2000 K, 2500 K, 300 K, 2000 K-NPT and geometry optimization data), and consisted of more than 200000 configurations from trajectories. After collecting the training data, in OUTCAR or vasprun.xml format, we transformed the atomic position information into local descriptors, thus reducing the dimensionality of the geometric information variables. In the choice of local structure descriptor, we used the two-atom embedding descriptor (se_e2_a). The se_e2_a was constructed from all location information for each atom, including angular and radial information of neighboring atoms.[50] The expected maximum number of each element was set to be C:96, H:96, N:64, Zn:16, which were determined by the number of each element in training set structures to make sure that the energy can be conserved and accuracy of the model is higher. The next step is to construct the structure-energy/force/virial mapping by reading the energy and force information from the VASP output file.

In this work, the deep-learning network architecture was configured with an embedding network containing three layers with neuron (25, 50, 100), along with a random seed included in each layer. Additionally, an ANN fitting network was employed, which consists of three layers with neurons (240, 240, 240) in each layer. The start and limit pre-factors for energy and force were set as follows: The energy start pre-factor is 0.02, and the energy limit pre-factor is 1. The force start pre-factor is 1000, and the force limit pre-factor is 1. The cutoff radius and cutoff radius smooth of each atom is set to 6.0 Å and 0.5 Å, respectively. More than one million (> 1000000) steps of interaction are used to train the deep potential function. For the structure of each set, we randomly select 70% of the configurations into the training set and 30% as the validation set. Finally, the DeePMD-kit code outputs a deep learning potential function file in the form of an atomic pair of potentials.[51]

**Deep Potential Molecular Dynamic (DPMD) in LAMMPS scheme**

All simulations were performed in LAMMPS in the isobaric-isothermal ensemble (NPT) with Nose-Hoover thermostat to equilibrate the system from 300 K to 2000 K at different heating velocities. The pressure is set to be zero with isotropic pressure in all directions. The timestep is set as 0.5 fs. The parameters for various simulation procedures, including their respective heating velocities, are given in Table 3.

**Dynamic Structural Analysis**

The radial distribution function was obtained by the statistical calculation of 50 snapshots in the equilibrium phase at the corresponding temperature. The thermodynamic quantities, including density and enthalpy, are set in LAMMPS to be output statistics every 200 steps (0.1 ps). The coordination number of N atoms surrounding the Zn atoms in ZIFs was computed using a cut-off radius of 2.5 Å. This specific value was chosen based on the Zn–N partial radial distribution function at room temperature.

**AUTHOR INFORMATION**

**Corresponding Author**

*Correspondence: lineng@whut.edu.cn

**Present Addresses**

†State Key Laboratory of Silicate Materials for Architectures, Wuhan University of Technology, Wuhan 430070, China

**Author Contributions**

N. Li conceived the project and designed all the calculations. Z. Shi conducted the calculations under N. Li supervised. Z. Shi, B. Liu, Arramel, Y. Yue, and N. Li analyzed the data and discussed the results. Z. Shi and N. Li wrote the manuscript. All authors reviewed and contributed to the final manuscript.

**ACKNOWLEDGEMENTS**

This work was supported by the National Natural Science Foundation of China (No.11604249); the Fok Ying-Tong Education Foundation for Young Teachers in the Higher Education Institutions of China (No. 161008); and Fundamental Research Funds for the Central Universities (No. WUT35401053-2022).